\documentclass[twoside,twocolumn,9pt]{article}
\usepackage{extsizes}
\usepackage[super,sort&compress,comma]{natbib} 
\usepackage[left=1.75cm, right=1.75cm, top=2cm, bottom=2.0cm]{geometry}
\usepackage{balance}

\usepackage{sectsty}
\allsectionsfont{\scshape}

\usepackage{graphicx} 
\usepackage{lastpage}
\usepackage{float}
\usepackage{fancyhdr}
\usepackage{fnpos}
\usepackage[english]{babel}
\usepackage{array}
\usepackage[T1]{fontenc}
\usepackage[usenames,dvipsnames]{xcolor}
\usepackage{setspace}
\usepackage{hyperref}
\usepackage[squaren]{SIunits}

\usepackage{upgreek}
\usepackage{xspace}
\usepackage{placeins}
\usepackage{amsmath,amssymb}

\begin{document}

\twocolumn[
  \begin{@twocolumnfalse}
\vspace{3cm}

\begin{center}

    \noindent\huge{\textbf{\textsc{Influence of the solid fraction on the clogging by bridging of suspensions in constricted channels}}} \\
    \vspace{1cm}

    \noindent\large{Nathan Vani,\textit{$^{a}$} Sacha Escudier,\textit{$^{a}$} Alban Sauret \textit{$^{a}$}}$^{\ast}$ \\

    \vspace{5mm}
    \noindent\large{\today} \\

    \vspace{1cm}
    \textbf{\textsc{Abstract}}
    \vspace{2mm}

\end{center}

\noindent\normalsize{Clogging can occur whenever a suspension of particles flows through a confined system. The formation of clogs is often correlated to a reduction in the cross-section of the channel. In this study, we consider the clogging by bridging, \textit{i.e.}, through the formation of a stable arch of particles at a constriction that hinders the transport of particles downstream of the clog. To characterize the role of the volume fraction of the suspension on the clogging dynamics, we study the flow of particulate suspensions through 3D-printed millifluidic devices. We systematically characterize the bridging of non-Brownian particles in a quasi-bidimensional system in which we directly visualize and track the particles as they flow and form arches at a constriction. We report the conditions for clogging by bridging when varying the constriction width to particle diameter ratio for different concentrations of the particles in suspension. We then discuss our results using a stochastic model to rationalize the influence of solid fraction on the probability of clogging. Understanding the mechanisms and conditions of clog formation is an important step for optimizing engineering design and developing more reliable dispensing systems.} \\

 \end{@twocolumnfalse} \vspace{0.6cm}

  ]

\makeatletter
\renewcommand*{\@makefnmark}{}
\footnotetext{\textit{$^{a}$~Department of Mechanical Engineering, University of California, Santa Barbara, California 93106, USA}}
\footnotetext{\textit{$^{*}$ asauret@ucsb.edu}}
\makeatother

\section{Introduction}

The flow of suspended particles can often lead to clogging in confined systems at various scales, from blood vessels\cite{patnaik1994vascular, ye2020key} and porous media,\cite{liu2019particle, bizmark2020multiscale} to pipes\cite{matsuo2009monitoring} and canyons.\cite{piton2022debris} Clog formation remains a major issue in many technological applications, whether it is in irrigation,\cite{adin1991dripper,cararo2006analysis,bounoua2016understanding,de2020clogging} filtration,\cite{japuntich1994experimental,song2006experimental} flood prevention,\cite{kia2017clogging} 3D printing,\cite{tlegenov2018nozzle,croom2021mechanics} biopharmaceutical manufacturing,\cite{giri2013prospects} or microfluidic devices.\cite{dressaire2017clogging,wyss2006mechanism,liot2018pore} Constrictions in channels are weak points for clogging in such systems since the cross-section of the channel suddenly reduces.\cite{dressaire2017clogging} Following the formation of a clog, further transport of particles downstream is prevented, potentially rendering the system useless and creating a growing filter cake.\cite{sauret2018growth,delouche2022flow} The subsequent increase in hydraulic resistance can either lead to a drastic decrease of the flow rate or an increase in pressure that can lead to sudden system failures.\cite{sauret2018growth, mokrane2020microstructure,souzy2022role}

 Three mechanisms can lead to the clogging of suspensions:\cite{dressaire2017clogging} (i) sieving in which a single entity is too large to pass a constriction,\cite{sauret2014clogging,duchene2020clogging} (ii) aggregation in which particles progressively deposit on the walls,\cite{agbangla2014collective,dersoir2015clogging,dersoir2019dynamics, delouche2022flow} and finally (iii) bridging in which several particles form a stable arch due to friction among themselves and with the channel walls.\cite{goldsztein2004suspension,marin2018clogging,hafez2021effect} In practical applications, these three mechanisms are often found to all be present and interacting. Particles can indeed form large aggregates able to sieve a constriction, or particles can aggregate on the walls, thus narrowing the channel and allowing for later bridging or sieving.\cite{dressaire2017clogging} A better understanding and modeling of clogging therefore requires identifying and modeling the different contributions to the clogging dynamics. Such efforts have been made for aggregation despite the complexity involved,\cite{wyss2006mechanism,dersoir2015clogging}  but relatively less for bridging. An improved understanding could help design new methods to prevent or delay clogging, \textit{e.g.}, as illustrated recently by the addition of pulsation to microfluidic flows.\cite{dincau2020pulsatile,dincau2022clog}

The clogging by arch formation, \textit{i.e.}, bridging, is usually observed for small to moderate width to particle diameter ratios and concentrated enough suspensions.\cite{valdes2006particle, roussel2007general, agbangla2012experimental} Some experimental studies have been done on suspension bridging at a constriction in microfluidic,\cite{sharp2005flow, marin2018clogging, souzy2020transition, hsu2021roughness} but also at larger scales. For instance, Guariguata \textit{et al.} considered disk floating at the air/liquid interface and reported the clogging probability of these particles at a constriction.\cite{guariguata2012jamming} Lafond \textit{et al.} have used a similar experimental system but with neutrally buoyant particles.\cite{lafond2013orifice} They characterized the backlog accumulation of a dilute suspension passing a small enough constriction, and the consequence of such an increase in solid fraction on the probability of clogging. In a microfluidic system, Marin \textit{et al.} have characterized the probability of clogging in a straight channel with an approximately square constriction of width $D = 100 \,{\rm \mu m}$ in a device of height $H = 1.1D$.\cite{marin2018clogging} They used particles of different diameters $d$ to probe the occurrence of clogging from a quasi two-dimensional situation to a three-dimensional one at a given volume fraction $\phi_{\rm V} \simeq 0.20 \pm 0.05$. They highlighted the important role of the ratio of the constriction size to the particle diameter, $D/d$, showing that in their configuration clogging can be observed only for $D/d <3$. In a following study, Souzy \textit{et al.} have considered the intermittent regime for volume fractions of the suspension close to the maximum packing. This regime is characterized by an alternating state of flowing and clogging and appears in their system for $2.43 < D/d < 5.26$.\cite{souzy2020transition} More recently, different studies have been exploring different physical parameters that can influence the clogging probability such as the role of the roughness,\cite{hsu2021roughness} and the influence of the deformability of the particles.\cite{bielinski2021squeezing} In addition, the difference between flow rate-driven flow and pressure driven-flow has recently been considered.\cite{souzy2022role}

While the number of studies considering the clogging by bridging of particles in suspensions is relatively limited, extensive research has been done on a related configuration: the flow and bridging of dry grains in silos. Several studies have considered the influence of the geometry of the silo either in two-dimensional\cite{zuriguel2011silo} or three-dimensional configurations.\cite{zuriguel2005jamming} Owing to the high level of control possible over the experimental system and its relative simplicity, the influence of various parameters on the clogging has been considered: the packing fraction in the range available for granular material under gravity,\cite{unac2012effect} the driving force by changing the gravitational field applied to the grains,\cite{dorbolo2013influence,arevalo2016clogging} the presence of an obstacle,\cite{zuriguel2011silo}, the addition of vibrations,\cite{lozano_breaking_2012, lozano_stability_2015} the velocity of the particles in the silo,\cite{huang2011influence,gella2017linking} and the limitation of the maximal flow rate of particles allowed.\cite{gella2018decoupling} Finally, a number of studies have been done based on the influence of the shape of particles,\cite{hafez2021effect,reddy2021clogging} and on their rigidity.\cite{hong2017clogging,tao2021soft}

However, there are important differences between the clogging by bridging of particulate suspension in constricted channels and the silo problem with dry granular systems. Some studies have focused on the comparison between the clogging dynamics of dry grains and gravity-driven immersed grains since the interstitial fluid can play a role.\cite{koivisto2017effect, hafez2021effect} While grains in silos are generally subject to gravity only, particulate suspensions can be driven by an imposed flow rate or by a pressure gradient.\cite{souzy2022role}  In addition, once the system has clogged, the energy is usually fully dissipated in silos, whereas the interstitial fluid still flows in the filter cake for particulate suspensions.\cite{sauret2018growth} The most striking difference between these systems is perhaps the role of the volume fraction on the clogging probability, which can vary significantly in suspension flows. Some numerical studies mentioned that bridging only occurs beyond a given volume fraction of particles for a given constriction size, whereas others did not find such a threshold.\cite{mondal2016coupled, sun2019investigating} Overall, quantifying the exact role of the volume fraction on clogging remains unclear despite the important fundamental and practical implications.

In this study, we use a quasi-bidimensional straight system made of a channel presenting a constriction of cross-section width $W$. We flow a suspension of particles at different solid fractions to probe its role in the clogging dynamics. We first describe the millifluidic channels used, as well as the suspensions of particles and the processing tools. We then report some examples of clogging events observed in our system. In particular, we introduce the relevant quantities used to describe the clogging: the average number of particles flowing through the constriction before clogging, the distribution of clogging events, and the evolution of these quantities with the solid fraction of the suspension and with the width of the channel. The results are then discussed and rationalized through a stochastic model to capture the evolution of the clogging probability with the solid fraction of the suspension. Our results suggest that the solid fraction of the suspension can be tuned to control the lifetime of a dispensing system but that for small values of $W/d$, the system will clog eventually.

\section{Experimental Methods} \label{sec:Experiments}

\begin{figure}
  \begin{center}
    \includegraphics[width=\linewidth]{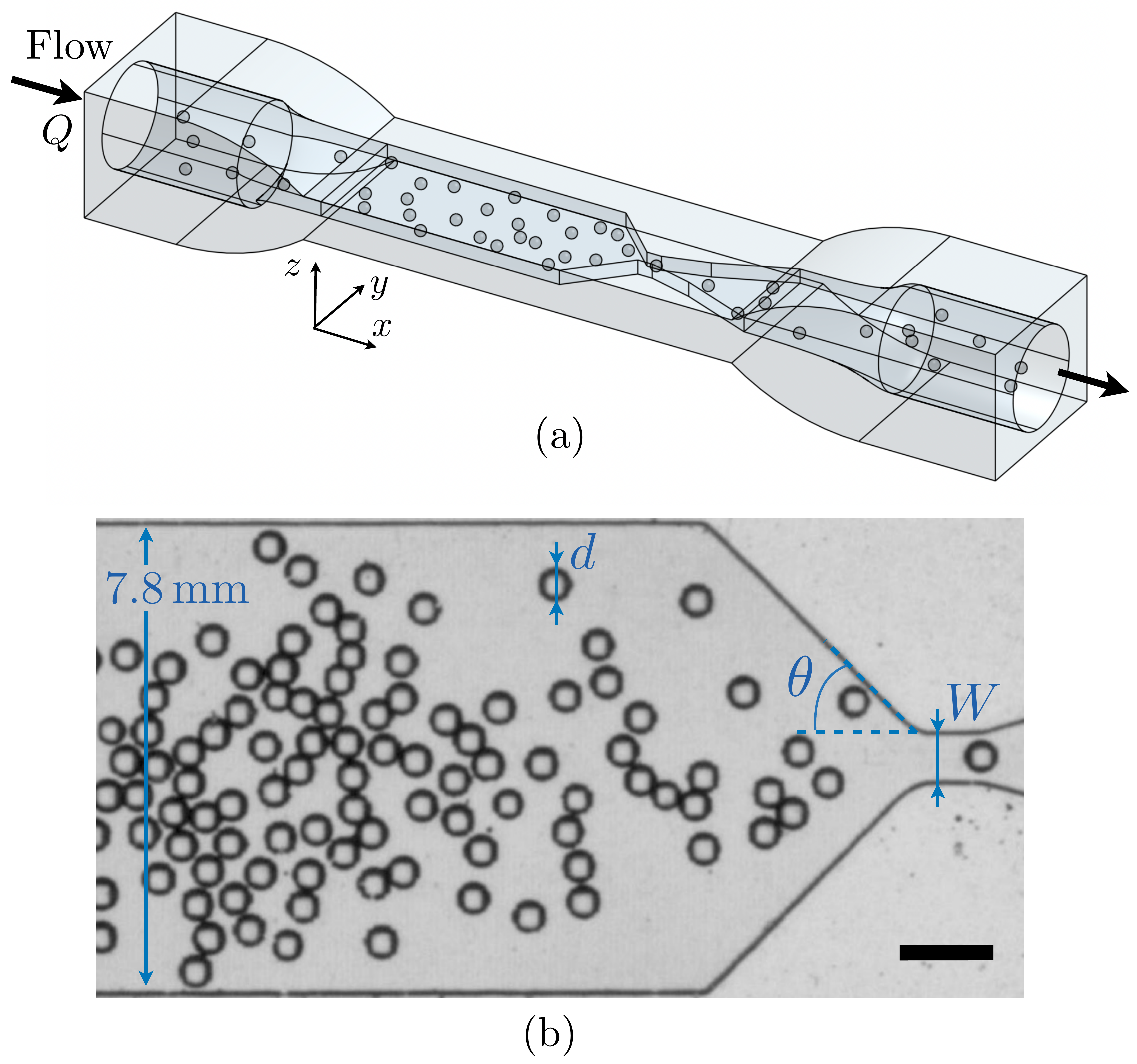}
  \end{center}
  \caption{(a) 3D computer-aided design (CAD) rendering and (b) picture of the millifluidic chip with particles flowing from left to right. A syringe pump is used to drive the flow of particles of diameter $d$ into the constriction of width $W$ and angle $\theta=45^{\rm o}$. The height of the channel is constant and equal to $H/d=1.6$. Scale bar is 3 mm.} 
  \label{fgr:figure1}
\end{figure}

The millifluidic channels are made using stereo-lithography (SLA) printing methods (Formlabs Form 3). This approach allows for a precise control of the geometry of the devices while keeping a quasi-bidimensional system. Indeed, past studies with dry grains in silos have shown that 2D and 3D geometries lead to different features.\cite{zuriguel2014invited} Therefore, in the present study, we restrict ourselves to a quasi-2D case. The geometry of the device is designed in a CAD software, and an example of rendering is shown in figure \ref{fgr:figure1}(a). The height of the channel is kept constant and equal to $H=0.95 \pm 0.02\,{\rm mm}$. We use the same particle diameter for all experiments presented here, $581 \pm 15 \,\mu{\rm m}$, so that the confinement ratio remains equal to $H/d \simeq 1.6$. Therefore, the particles do not fully overlap, allowing their precise tracking throughout the device. The width of the channel is $7.8\,{\rm mm}$ and the channel is narrowed linearly down to the constriction width with an angle of $\theta=45^{\rm o}$. A picture of the device and the notations used are shown in figure \ref{fgr:figure1}(b). Several channels are used with different constriction widths $W$, while keeping all the other geometrical parameters of the system constant. In the present study, we consider widths in the range  $0.87\,{\rm mm} \leq W \leq 2.40\,{\rm mm}$, corresponding to $1.5 \leq W/d \leq 4.2$.

The suspensions are made of neutrally buoyant and non-Brownian spherical particles dispersed in a Newtonian fluid (see supplementary information for more details). The particles used are polystyrene beads (DynoSeeds from MicroBeads), with a diameter of $581 \pm 15 \,\mu{\rm m}$ and density $\rho_{\rm p} \simeq 1.049 \,{\rm g.cm^{-3}}$. They are dispersed in a Newtonian fluid consisting of a mixture of water and polyethylene glycol (PEG) at $38\%/62\%$ per weight. The measured viscosity of the interstitial fluid is $\eta_{\rm f}=75\,{\rm mPa\,s}$ and the density of the fluid, $\rho_{\rm p}=1.049 \,{\rm g.cm^{-3}}$, closely matches the density of the particles. The electrostatic forces are negligible due to the relatively large size of the particles, which prevents their their agglomeration or their deposition on the wall of the channel. As a result, only clogging by bridging, \textit{i.e.}, through the formation of arches, is present in our experiments.

The suspension is injected into the millifluidic device at a given volume fraction $\phi_{\rm V}$ using a syringe pump (KDS Legato 110) at a flow rate $Q=1\,{\rm mL.min^{-1}}$ leading to a typical velocity in the main channel of $u_{\rm p}=2.4\,{\rm mm.s^{-1}}$. The resulting particle Reynolds number, ${\rm Re_p}={\rho_f\,d\,u_p}/{\eta_f}$ is of order $0.02$ in the channel, and in the range $0.06$ to $0.15$ at the constriction. We also performed some experiments at other flow rates but ensured that ${\rm Re_{\rm p}}$ remains smaller than 1. The outlet of the system is at ambient pressure so that the pressure ahead of the constriction remains constant.\cite{sauret2018growth}

Since our experiments are quasi-bidimensional ($H/d \simeq 1.6$), and following approaches done in 2D silos, we consider in the following the surface fraction of particles defined as $\phi=A_{\rm p}/A_{\rm tot}$, where $A_{\rm p}$ is the projected surface area of the particles and $A_{\rm tot}$ the total surface area. The volume fraction $\phi_{\rm V}$ and the surface fraction of particles are related through $\phi_{\rm V}/\phi = 2\,D/(3\, H)$. However, due to the tubing connectors, the syringe, and possible self-filtration mechanism prior to the millifluidic chip,\cite{haw2004jamming} the measured surface fraction can be slightly smaller than the expected value. As a result, we initially performed calibration tests in a straight channel with no constrictions to measure the resulting surface fraction for different input volume fractions (see supplementary information). We also measured the surface fraction in the channel before the constriction for each experiment.

\begin{figure*}[h]
  \begin{center}
    \includegraphics[width=\linewidth]{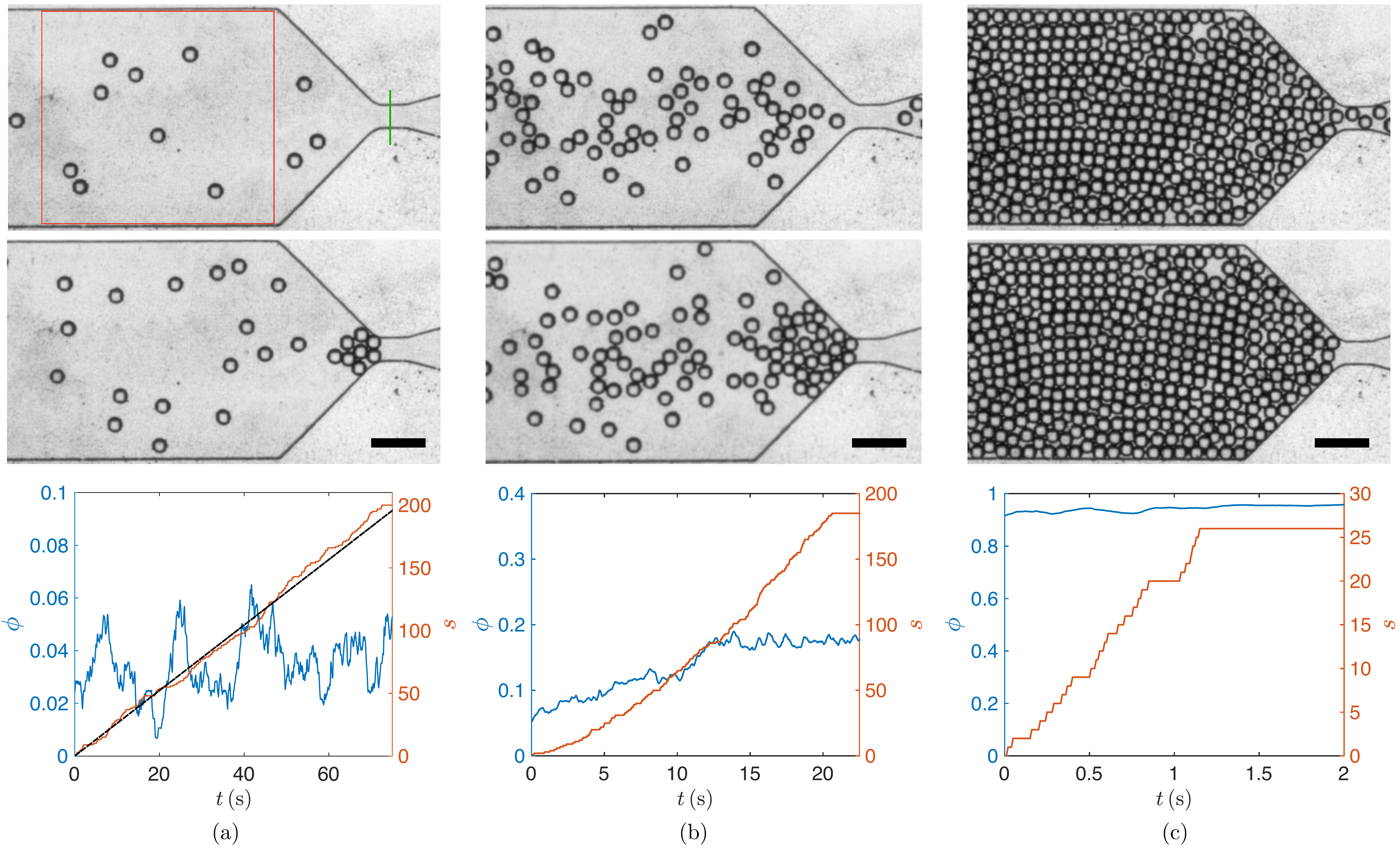}
  \end{center}
  \caption{Snapshots of the particles flowing in the channel before (top images) and after (middle images) a clogging event for a surface fraction (a) $\phi \simeq 0.03$, (b) $\phi \simeq 0.17$, and (c) $\phi \simeq 0.95$ for a constriction $W/d=1.7$. The bottom plots show the instantaneous surface fraction $\phi(t)$ (blue line) measured in the red box shown in (a) and the time evolution of the cumulative number of escapees $s$ (orange line) measured at the green line shown in (a). The dashed line in (a) corresponds to the prediction of the time-evolution of the number of escapees before clogging given by Eq. (\ref{eq:new_proof}). The scale bars are 2 mm.} 
  \label{fgr:figure2}
\end{figure*}

The flow of particles in the channel is recorded using a camera (Edmund Optics, 2323M) at $100\,{\rm fps}$ with a resolution of $60\,\mu{\rm m}/{\rm pix}$ and the system is backlit with a LED Panel (Phlox). The videos are then processed with a custom-made Python routine. The positions of the particles at each time frame are extracted, and their trajectories are analyzed using Trackpy.\cite{allan_daniel_b_2021_4682814} This approach allows us to determine individual particle trajectories, their velocities, the instantaneous surface fraction $\phi$ in the channel, and the time-evolution of the number of escapees at the constriction $s$. We can then detect the formation of a clog and determine the size of the arch. For each experimental parameter, a sufficiently large number of trials are performed to obtain reliable statistical data on the clogging dynamics. After a clogging event has occurred, the system is reset by reversing the flow until no particle remains in the device or upstream in the system. The suspension that previously passed through the device is discarded through a side tubing, and a new batch of suspension at the same solid fraction is used for the next clogging event when the flow is re-established. We ensured that there is no history in the clogging events by ensuring that there is no correlation between a clogging event and the following one (see supplementary information).

\section{Experimental Results}  \label{sec:Results}

\subsection{Phenomenology}

Examples of the clogging dynamics for three different surface fractions of particles $\phi$ and a particle to constriction size ratio of $W/d=1.7$ are shown in figures \ref{fgr:figure2}(a)-(c). We also report the time-evolution of the surface fraction $\phi(t)$ upstream of the constriction (shown by the red box in figure \ref{fgr:figure2}(a)). Finally, in all experiments, we count the number $s$ of particles flowing through the constriction before a clog forms.

The closest configuration to the classical dry granular clogging is for $\phi$ close to the maximum packing fraction, as reported in Fig. \ref{fgr:figure2}(c). In this situation, owing to the relatively small value of $W/d$, the clogging occurs quite quickly, with only $26$ particles escaping before forming a stable arch in the case shown here. The particle surface fraction remains stable during the entire experiment and equals to $\phi=0.95 \pm 0.02$ here. Note that the large surface fraction value comes from the fact that the system is not perfectly bidimensional since $H/D = 1.6$.

For very dilute suspension, $\phi=0.035 \pm 0.01$ in figure \ref{fgr:figure2}(a), the flow is uninterrupted for a long time. Since the suspension is much more dilute, the number of particles flowing through the constriction $s$ does not increase as fast over time compared to the dense case. Still, a clog eventually occurs, here at $t=75\,{\rm s}$ (compared to $t=1.2\,{\rm s}$ in the dense case). The total number of escaped particles, $s=200$, is also larger than for the dense case mentioned previously. We should emphasize that the choice of the control volume to estimate the solid fraction, providing that the control volume is large compared to the particle size, does not influence the value of $\phi$. Before clogging and once the flow of particles is fully established, we can relate the volume fraction and the number of particles flowing through the constriction:
\begin{equation}\label{eq:new_proof}
s(t)=\frac{6\,\phi_{\rm V}\,Q\,t}{\pi\,{d}^3}.
\end{equation}

Finally, we show in figure \ref{fgr:figure2}(b) a case with an intermediate surface fraction, $\phi=0.16 \pm 0.03$ before clogging. Overall, the dynamics is similar, but the surface fraction shows more variation around the mean value than the dense case due to the discrete nature of particles. In addition, the experimental method requires to start with pure fluid in the device before flowing the suspension so that the particle solid fraction needs to quickly ramp up to its target value. Again, the constriction eventually clogs after a given number of particles flow through.

In our experimental systems, clogging events have been observed for higher values of $W/D$, for dense cases up to $W/D=4.2$. Nonetheless, experimental limitations to obtain a reliable statistical analysis prevent the extensive study of those cases. We should also emphasize that some intermittency, \textit{i.e.,} the formation of metastable arches breaking after a given amount of time, can be observed in our system for $W/D \gtrsim 3.7$ at the largest possible solid fraction ($\phi=0.95$). The exact dynamics of this intermittency is beyond the scope of this study but has been characterized recently in 3D for dense suspensions.\cite{souzy2020transition} In the following, we define a clogging event as the formation of an arch lasting for longer than $\tau = 5\,d/u_{\rm p} > 1.5 {\rm s}$, where $d/u_{\rm p}$ is the Stokes time associated with the particles.\cite{souzy2022role} 

In summary, these different examples show that even when diluting the suspension significantly, there still exists a probability of clogging the constriction. Experiments with even more dilute suspensions have been made, showing that clogging events can happen even with a solid fraction of around $\phi=0.01$ for small constrictions ($W/d<2$) but become very rare ($s > 1000$). Experimental limitations prevent us from probing those cases extensively. Nevertheless, a situation in which a sufficient number of particles being close enough and well-placed to form an arch remains a possibility even in very dilute cases, as a flawless uniform distribution of particles is not expected in particulate suspensions. Indeed, particles flowing in a channel experience hydrodynamic interactions between themselves. As a result, they can form temporary non-adhesive pair (or more) of particles,\cite{haddadi2015topology,schaaf2019flowing,fouxon2020theory} which is favorable to clogging by bridging. The existence of such transient assembly of particles does not depend on the volume fraction, although their frequency of formation is. We should emphasize that even if conceptually the clogging of a small enough constriction appears to only be a matter of time for any solid fraction, the timescale over which clogging would occur for very dilute suspensions (\textit{e.g.}, $\phi=10^{-4}$) would likely result in a system that does not clog in practical applications over a limited time.

In the following, we consider quantitatively the influence of the solid fraction on the clogging dynamics, and in particular, we focus on the number of particles escaping the constriction before clogging.  This metric can also be seen as the volume of suspension that can be dispensed through the system before it clogs.

\subsection{Distribution of particles flowing through the constriction before clogging}

\begin{figure*}[ht]
  \begin{center}
    \includegraphics[width=\linewidth]{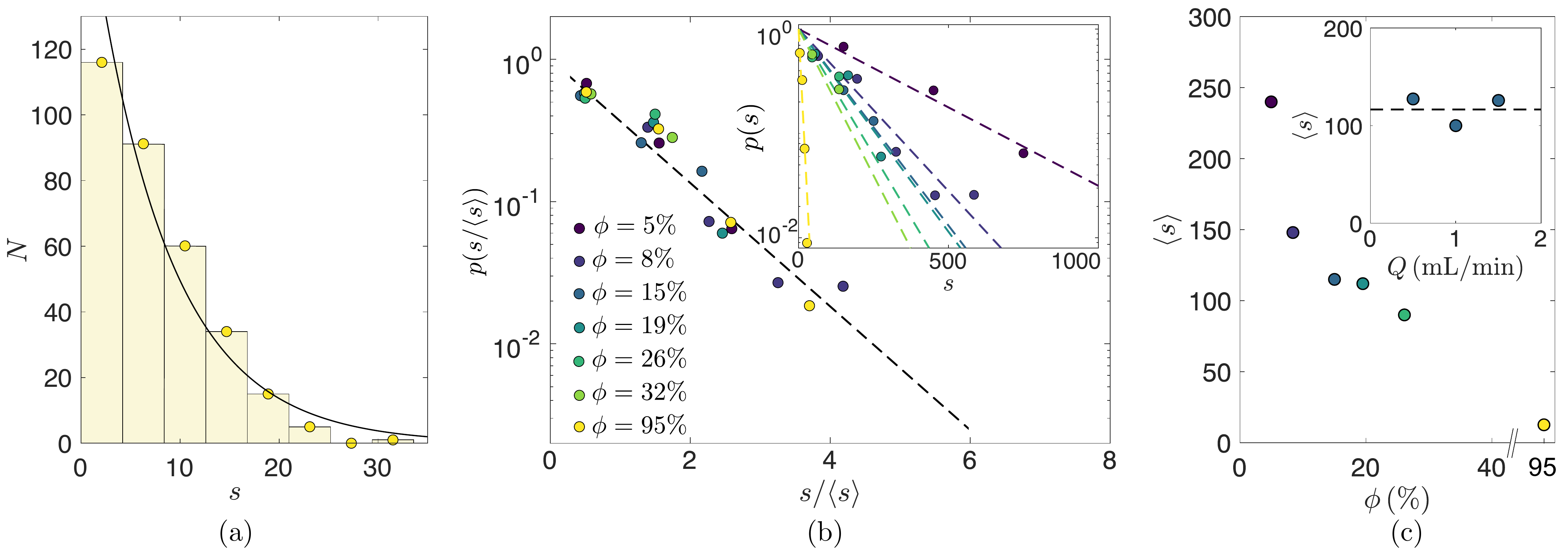}
  \end{center}
  \caption{(a) Example of a histogram showing the number of events $N$ for which $s$ particles flow through the constriction before clogging, with $W/d=1.7$ and $\phi=0.95$. The circles are the data extracted and reported in (b) and the solid line is an exponential distribution. (b) PDF $p(s/\langle s \rangle)$ of the normalized number of particles escaping the constriction $s/\langle s \rangle$ before clogging. Inset: PDF $p(s)$ of the number of particles escaping the constriction $s$ for different surface fraction of particles. The dotted lines show exponential fits and the different colors indicate different surface fraction of particles $\phi$. (c) Evolution of the mean number of particles flowing through the constriction before clogging $\langle s \rangle$ for various particle surface fraction $\phi$ (same color code than in (b)). Inset: mean number of particles escaping the constriction, $\langle s \rangle$, when varying the flow rate for $\phi=0.15$. The dotted line shows the average value. All experiments reported here are for a constriction $W/d=1.7$.}
  \label{fgr:figure3}
\end{figure*}

There are different ways to describe clogging in constricted systems. Some studies report a clogging/non-clogging diagram for a volume of suspensions dispensed.\cite{valdes2006particle, hsu2021roughness, bielinski2021squeezing} Other studies report a clogging probability by performing a given number of experiments or numerical simulations.\cite{mondal2016coupled,hong2017clogging,sun2019investigating,hafez2021effect} Finally, an approach commonly used for the clogging of silos is to count the number of particles that pass through the constriction before a clog occurs, \textit{i.e.}, counting the number of escapees.\cite{zuriguel2005jamming,saraf2011power, lafond2013orifice, marin2018clogging} In other words, the goal of this approach is to characterize the avalanche size distribution after performing a statistically relevant number of trials. In the case of dry granular materials, it has been shown that the avalanche size distribution follows an exponential decay in 2D and 3D configurations.\cite{janda2008jamming,zuriguel2014invited} A similar behavior has been reported recently for a volume fraction $\phi_{\rm V} \simeq 0.2$ in the case of a non-Brownian suspension in a constricted channel,\cite{marin2018clogging} as well as for suspensions close to the packing fraction, but where the system goes from 2D to 3D.\cite{souzy2020transition} This exponential decay of the distribution of escaping particles shows that the probability for a particle to pass through the constriction without clogging is constant throughout the event. Since clogging events are uncorrelated (see supplementary information), every particle has the same probability of clogging the constriction. We can therefore consider that the probability $p_{\rm clog}$ for a given particle to clog the system is independent from that of other particles. The avalanche size distribution can thus be modeled with a geometric distribution, with the probability of observing $s$ particles escaping before the formation of a clog being $p(s)= p_{\rm clog}\left(1- p_{\rm clog} \right)^s$.\cite{zuriguel2005jamming,zuriguel2014invited}

For one trial, the number of escapees $s$ corresponds to the final value of $s(t)$, as shown for instance in figure \ref{fgr:figure2}(c) where $s=26$. We perform a large number of experiments and report in figure \ref{fgr:figure3}(a) an example of the histogram of the number $N$ of clogging events having a number $s$ of escapees for $W/d=1.7$ and $\phi=0.95$. As expected, $N$ decreases with increasing $s$ and the resulting histogram can be fitted by an exponential curve, which is characteristic of a geometric distribution. This exponential tail is observed for all solid fractions considered here, as shown in the inset of Fig. \ref{fgr:figure3}(b). We can also calculate the average number of particles escaping the constriction before clogging, $\langle s \rangle $, and rescale the probability distribution function (pdf) of the number of escapees $p(s/\langle s \rangle)$. In this case, all the distribution collapse on a master curve exhibiting an exponential decay as shown in \ref{fgr:figure3}(b). Therefore, we can describe the clogging by bridging of suspensions with a framework similar to the one developed for dry grains in silos.\cite{zuriguel2014invited} Consequently, to capture the clogging of suspensions by bridging in a constricted channel only requires to predict the average number of escapees before clogging, $\langle s \rangle $, when varying both the surface fraction of the suspension and the width of the constriction.

\subsection{Average number of escapees before clogging}

We report in figure \ref{fgr:figure3}(c) the evolution of the average number of particles $\langle s \rangle $ flowing through the constriction of width $W/d=1.7$ before a clogging event occurs when varying the particle surface fraction $\phi$. Qualitatively, $\langle s \rangle $ increases when decreasing $\phi$, meaning that the probability of clogging decreases when less concentrated suspensions are flowing through the constriction as one could expect. Interestingly, even at relatively low surface fractions ($\phi <0.1$), the particles are still able to clog the constriction. This result is in disagreement with results from numerical simulations that suggested the existence of a ``critical particle concentration'' for spontaneous bridging to occur.\cite{mondal2016coupled} In the present case, and similarly to the study of Marin \textit{et al.},\cite{marin2018clogging} our experiments suggest that even if it will take a longer time for the system to clog, eventually, there is a non-zero probability for a small enough constriction to be clogged at some point. 

We also perform some experiments for $W/d=1.7$ and $\phi\simeq 0.15$ varying the flow rate of the suspension in the range $0.5 \,{\rm mL/min} \leq Q  \leq 1.5 \,{\rm mL/min} $, corresponding to particle Reynolds number before the constriction of ${9 \times 10^{-3} \leq  \rm Re_{\rm p} \leq 3 \times 10^{-2} }$. The evolution of $\langle s \rangle$ is shown in the inset of figure \ref{fgr:figure3}(c). For this range of small Reynolds numbers, we do not observe a significant or monotonic evolution of $\langle s \rangle$ in our experiments. Therefore, in the following we perform all experiments at a given flow rate $Q=1\,{\rm mL/min}$. Nonetheless we should emphasize that, at larger particle Reynolds numbers, inertial effects might have an effect on the clogging dynamics.

\section{Discussion}   \label{sec:discussion}

To describe the influence of the volume fraction on the clogging probability, we consider a suspension of particles of diameter $d$, which do not modify the flow and passively follow the streamlines since the particles are neutrally buoyant. In addition, we consider that the particles are initially randomly distributed in the channel with a uniform probability distribution in space. The experiments are performed at a constant flow rate so that the volume of suspension flowing through the constriction is constant over time. Our experiments are quasi-bidimensional, so that we use the surface fraction of particles $\phi$, and we consider a small surface area $\Sigma$ of suspension that flows in the channel until it reaches the constriction of width $W$.  Since the particles are assumed to be randomly distributed in the channel, the number of particles that reach the constriction for a given time interval also follows a random distribution. 

For a clogging event to occur, three conditions must be fulfilled simultaneously. First, (i) the formation of an arch of $n=\lvert W/d \rvert+1$ particles, meaning that the arch must span the entire constriction. This requires that (ii) when one particle arrives at the constriction, $n-1$ other particles must be contained in the area $\Sigma= (n-1)\,\pi\,d^2/(4\,\phi_{max})$, corresponding to a dense packing of particles [see top inset of figure \ref{fgr:figure4}]. Finally, once an arch is formed, (iii) the particles must be in a configuration such that the arch is stable. Indeed, the particles may form an arch that is in an unstable configuration and which immediately breaks after being formed, thus not leading to a clogging event.\cite{to2001jamming} We attribute a probability $\alpha(n)$ to this occurrence. The expression of $\alpha$ depends on the number of particles $n$ forming the arch. Since here we only change the surface fraction and the width of the constriction, we consider the ansatz $\alpha(n)={\alpha_0}^{n-1}$, where $\alpha_0$ is a constant in our case but in more general situations may depend on different experimental parameters such as the angle of the constriction, the friction coefficient between the particles, the nature of the flow, and other experimental parameters.

\subsection{Discrete stochastic model}

We first rely on a discrete stochastic model following the approach developed in Refs. [29] and [68] \cite{goldsztein2004suspension,goldsztein2005volume}. To do so, we generate numerically a random distribution of particles with a surface fraction $\phi$ in the channel. Note that the particles can overlap in this model since we only consider the position of their center of mass. Then, we compute numerically for each particle arriving at the constriction the number of particles in the adjacent area $\Sigma$ to fulfill conditions (i) and (ii). When these conditions are satisfied, an arch is formed. We now add the stochastic condition (iii) that the arch must be stable. If all conditions are fulfilled, we consider that the channel is clogged. Otherwise, we move on to the next particle arriving at the constriction. We can then calculate the total number of particles flowing through the constriction before clogging, $s$, by summing all the particles that did not lead to clogging before the conditions (i) to (iii) were all satisfied. 

\begin{figure}[h!]
  \begin{center}
 \includegraphics[width=\linewidth]{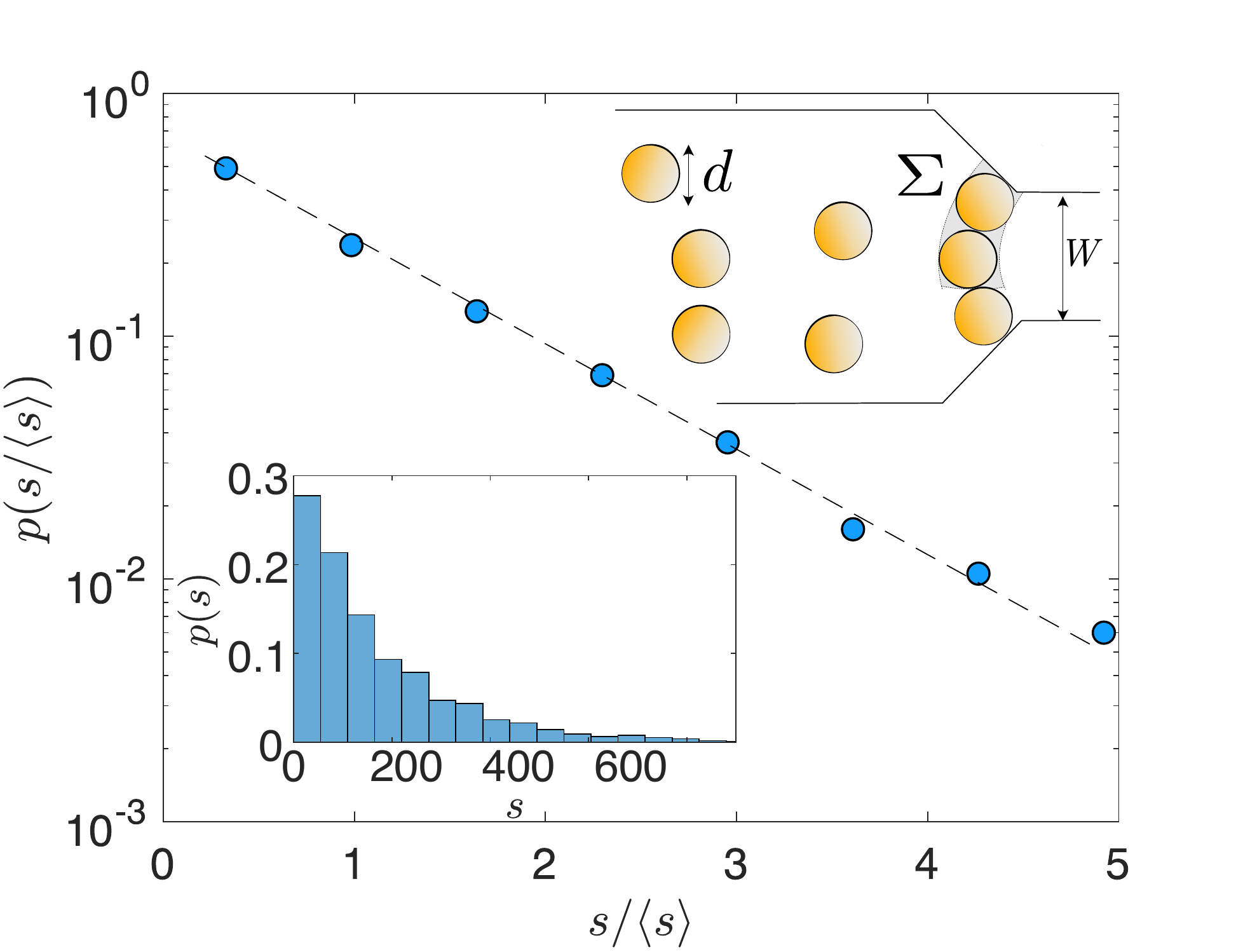}
  \end{center}
  \caption{Example pdf $p(s/\langle s \rangle)$ of the normalized number of particles escaping the constriction $s/\langle s \rangle$ before clogging obtained with the stochastic model. The dashed line is an exponential law. Top inset: Schematics of the situation considered in the model. The area that must contains $n-1$ particles, $\Sigma= (n-1)\,\pi\,d^2/(4\,\phi_{max})$, for an arch to form when one particle arrives at the constriction is shown in grey. Bottom Inset: Example of histogram of the number of escaped particles $s$. This simulations are performed with $W/d = 2.8$ and $\phi=0.2$.}   \label{fgr:figure4}
\end{figure}

An example of a result of this discrete stochastic model is shown in figure \ref{fgr:figure4} for a surface fraction $\phi=0.2$ and a constriction $W/d=2.8$. We observe that the histogram of escapees $s$ before clogging shows an exponential tail [bottom inset of figure \ref{fgr:figure4}], as expected from the fact that the formation of arches at the constriction is an uncorrelated process. We plot in the main panel of figure \ref{fgr:figure4} the evolution of the number of particles flowing through the constriction before clogging rescaled by the mean number of escapees. The dashed line indicates the exponential behavior. 

The stochastic model presents one fitting coefficient, $\alpha_0$, which is estimated with our experimental results. For the suspension used here and an angle of the constriction $\theta=45^{\rm o}$, we find that $\alpha={\alpha_0}^{n-1}$ with $\alpha_0=0.08$ captures our experimental results well, as can be seen in figures \ref{fgr:figure5}(a)-(b). A thorough understanding of the evolution of $\alpha(n)$ with the different experimental parameters of the problem is beyond the scope of the present study but would provide an interesting predictive tool for the clogging in constricted flow. One may expect, for instance, that $\alpha_0$ will increase with the roughness of the particles.\cite{hsu2021roughness}

From the numerical computation of the stochastic model, we can extract the average number of escapees $\langle s \rangle$ when varying the surface fraction and for different values of $W/d$, as reported in figures \ref{fgr:figure5}(a)-(b). We recover the decrease of the average number of particles flowing through the constriction $\langle s \rangle$ when increasing the surface fraction, as observed experimentally in figure \ref{fgr:figure3}(c). In addition, the rescaled width of the constriction $W/d$ has a very strong influence.\cite{marin2018clogging} Note that the model predicts jumps in the number of particles escaping before clogging when $n = \lfloor {W}/{D} \rfloor + 1$ increases by one. Indeed, one additional particle is needed to form an arch. However, qualitatively one could also expect that an arch made of $n=3$ particles spanning a constriction of $W/d=2.3$ has more potential stable configurations than an arch of $n=3$ particles for constriction of $W/d=2.9$.\cite{to2001jamming} Therefore, other possible ansatz for $\alpha(n)$ that would introduce a weak dependence on a step of constant $n$ could be considered in the future, but here our expression is sufficient to capture the key features.

\subsection{Modeling the average number of escapees}

We can also consider the relevance of an analytical approach to capture the evolution of $\langle s \rangle$.\cite{marin2018clogging} We introduce the total surface in the channel $\mathcal{A}$, and the corresponding total number of particles $N_{\rm part}$ in this area, so that the surface fraction of particles is  given by $\phi=\pi\,N_{\rm part}\,d^2/(4\,\mathcal{A})$. The conditions (i) and (ii) are given by the probability of finding $n-1$ particles in the area $\Sigma$ (with $k < N_{\rm part}$ and $\Sigma < \mathcal{A}$):
\begin{equation} \label{eq:proba_demo}
    p_1 =  \frac{N_{\rm part}!}{(n-1)!\,(N_{\rm part}-n+1)!} \left( \frac{\Sigma}{\mathcal{A}}\right)^{n-1} \left( 1 - \frac{\Sigma}{\mathcal{A}} \right)^{N_{\rm part}-n+1}.
\end{equation}
In the limit $N_{\rm part}\gg n$, we can write
\begin{equation}
    \frac{N_{\rm part}!}{(n-1)!\,(N_{\rm part}-n+1)!} \xrightarrow[N_{\rm part}\gg k]{} \frac{{N_{\rm part}}^{n-1}}{(n-1)!}.
\end{equation}
In addition, since $\Sigma \ll \mathcal{A}$ the last term of the expression becomes
\begin{equation}
\left( 1 - \frac{\Sigma}{\mathcal{A}} \right)^{N_{\rm part}-k} \xrightarrow[N_{\rm part}\gg k]{\Sigma \ll \mathcal{A}} {\rm exp}\left(-\frac{4\,\phi\,{\Sigma}}{\pi\,d^2}\right).
\end{equation}
Substituting the expression of $\Sigma$ in equation (\ref{eq:proba_demo}), we obtain the probability of forming an arch of $n$ particles (conditions (i) and (ii)):
\begin{equation}
p_1=\frac{1}{(n-1)!}\,\left[\frac{(n-1)\,\phi}{\phi_{\rm m}}\right]^{n-1}\,{\rm e}^{-(n-1)\,\phi/\phi_{\rm m}}.
\end{equation}
We now also need the condition (iii) for the stability of the arch formed. We consider that the probability of the arch made of $n$ particles to be stable is
\begin{equation}
p_2(n)=\alpha(n)={\alpha_0}^{n-1}.
\end{equation}
This ansatz captures that more particles lead to less stable arch,  satisfies the limit $p_2(1)=1$ (since only one particle forms the clog) and that at large $n$, $p_2 \to 0$. This expression also captures the jump in probability when one additional particle is added to the arch.

Since the two conditions are independent, the probability of forming a stable arch is thus given by:
\begin{align}
p_{\rm clog}(\phi,n) &= p_1(\phi,n)\,p_2(n)\\
       &=\frac{{\alpha_0}^{n-1}}{(n-1)!}\,\left[\frac{(n-1)\,\phi}{\phi_{\rm m}}\right]^{n-1}\,{\rm e}^{-(n-1)\,\phi/\phi_{\rm m}}.
\end{align}

In the system considered here, the clogging events follow a geometric distribution meaning that the probability of clogging $p_{\rm clog}$ is the same after $s$ particles have flowed through the constriction. The probability of clogging after $k$ particles escape the constriction is thus $p(s=k)=p_{\rm clog}\,(1-p_{\rm clog})^k$ so that one could also write this probability in a continuous analogue using an exponential distribution. The average of this leads to the expected value for the number of particles flowing through the constriction before the first clog appears:
\begin{equation}\label{eq:smean}
    \langle s \rangle = \frac{1-p_{\rm clog}}{p_{\rm clog}}=\frac{\displaystyle (n-1)!}{\left[{\alpha_0\,(n-1)\,\phi}/{\phi_{\rm m}}\right]^{n-1}}\,{\rm e}^{(n-1)\,\phi/\phi_{\rm m}}-1
\end{equation}

\begin{figure}[h!]
  \begin{center}
 \includegraphics[width=\linewidth]{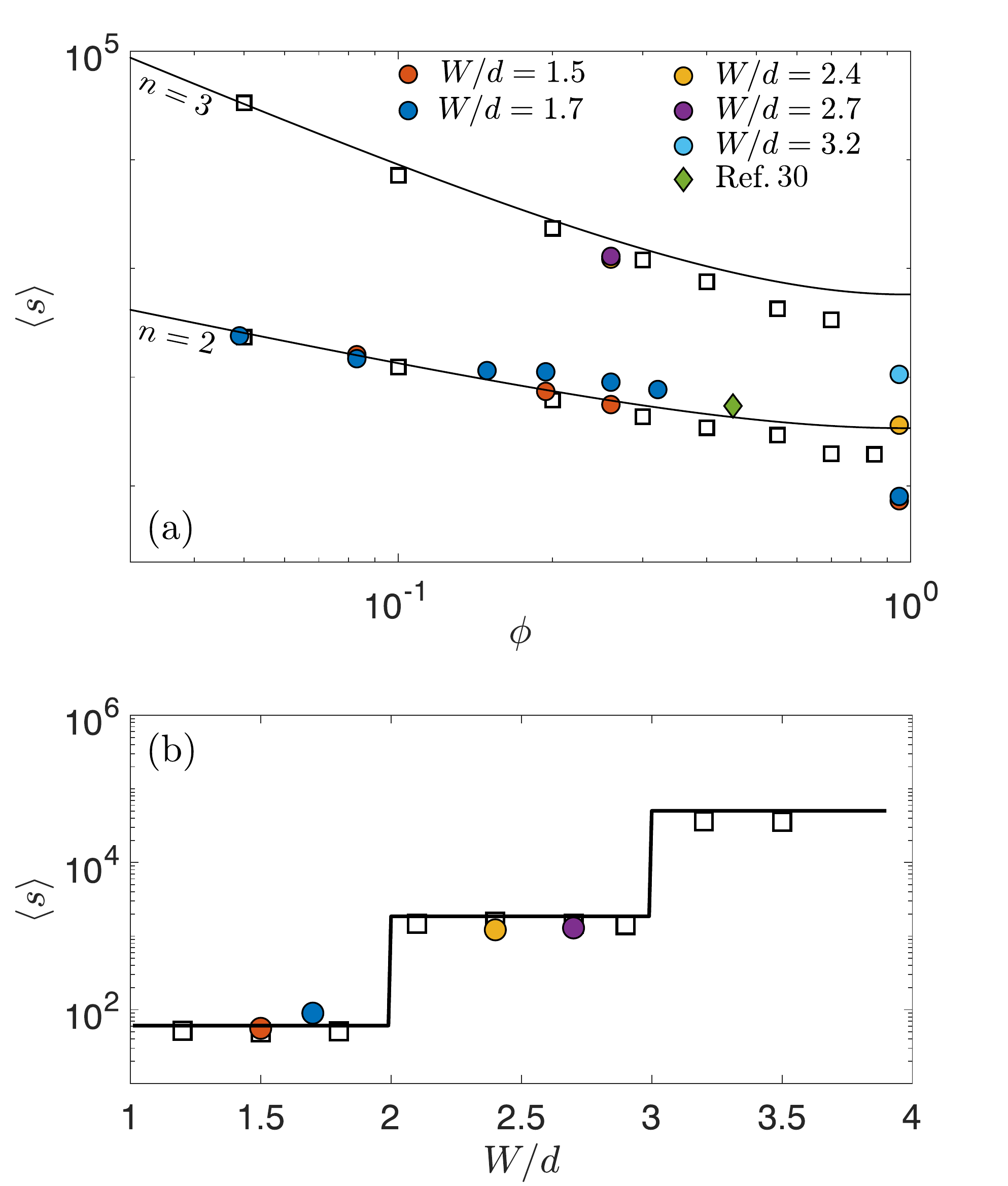}
  \end{center}
  \caption{(a) Evolution of the mean number of escapees $\langle s \rangle$ with the surface fraction of particles $\phi$. The color circles are the experimental points obtained for various values of $W/d$ (see legend), the green diamond is an experimental data from Marin \textit{et al.} \cite{marin2018clogging}, and the black squares are the result of the discrete numerical model. The solid lines are the prediction of the model given by Eq. (\ref{eq:smean}) with $\alpha_{0}=0.08$ for different values of $n$. (b) Evolution of the number of escapees with the dimensionless width of the constriction $W/d$ for $\phi \simeq 0.26$. The line shows the prediction for the model with $\alpha_{0}=0.08$. The symbols are the same than in (a).}   \label{fgr:figure5}
\end{figure}

We report equation (\ref{eq:smean}) in figure \ref{fgr:figure5} using the same value of $\alpha_0=0.08$ than the one used in the discrete stochastic simulation. Equation (\ref{eq:smean}) provides an analytical expression that efficiently captures the experiments and the discrete simulations for moderate surface fractions of particles, \textit{i.e.}, when the assumptions of the model are valid, and with the same trends when varying $\phi$ and $W/d$.

The model shows that decreasing $\phi$ for a given value of $W/d$ increases by orders of magnitude the number of particles flowing through the constriction before clogging occurs. Interestingly, using the experimental data point of Marin \textit{et al.}\cite{marin2018clogging} for the condition in which their system is also quasi-2D ($D/d=1.7$) and extrapolating their volume fraction to a surface fraction $\phi$ also leads to a good agreement with our data despite the differences between the two experimental systems (see green square in figure \ref{fgr:figure5}). 

We should emphasize that the model presents some limitations: the particles are randomly distributed and can overlap. Therefore, such an approach is mainly valid for dilute and moderate solid fractions. Besides, the evolution of the probability of stability of an arch, ${\alpha_0}^{n-1}$, accounts for the complexity of the system and is going to depend on different experimental parameters. This ansatz leads to a good description of our system and is correlated to the probability of forming a stable arch once a sufficient number of particles arrive at the constriction. However, this ansatz is expected to be different in 3D and likely depends on the roughness of the particles, the angle of the constrictions, as well as the fluid flow velocity.

\subsection{ Peculiar case of dense suspensions}

\begin{figure}[ht]
  \begin{center}
    \includegraphics[width=\linewidth]{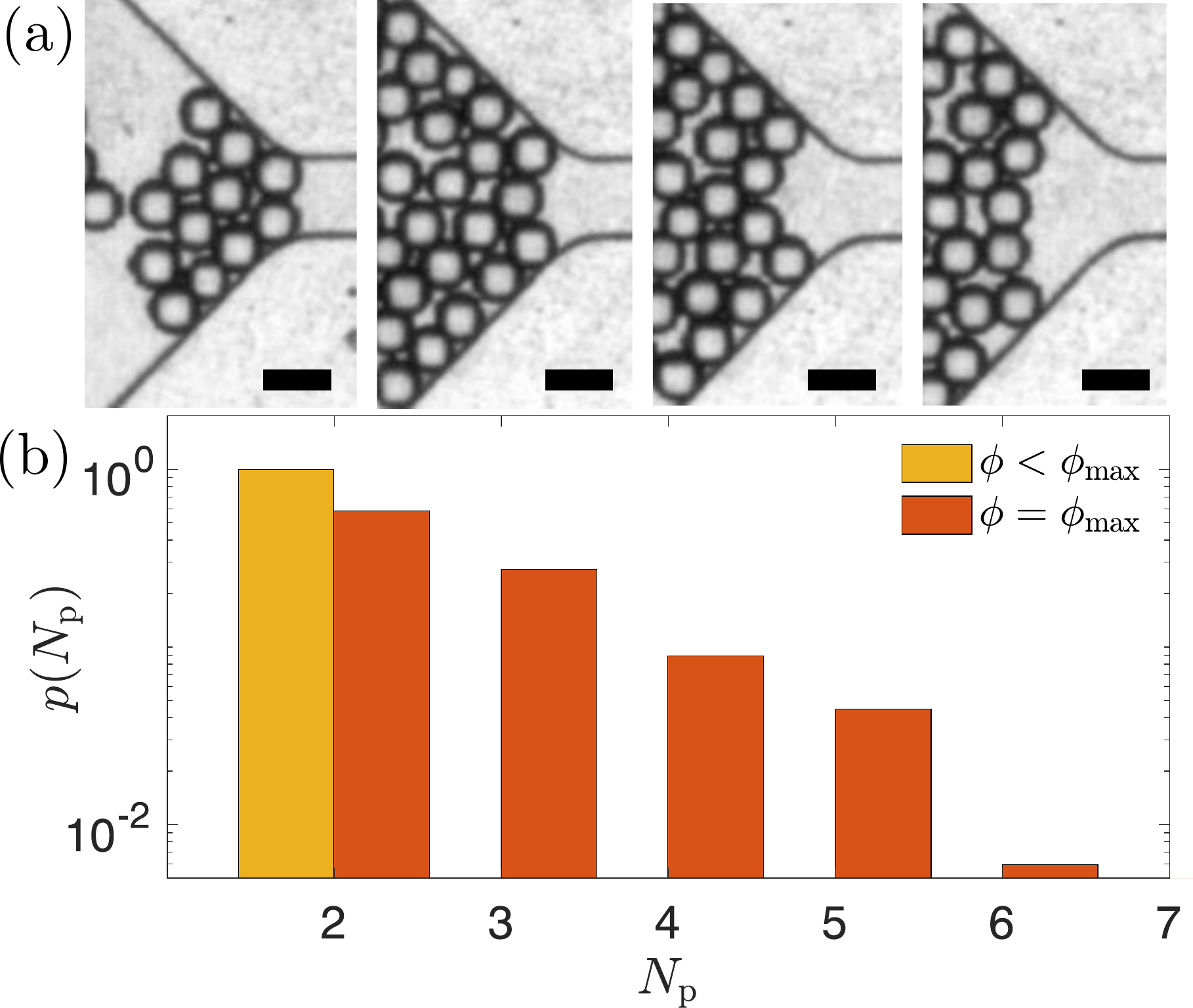}
  \end{center}
  \caption{(a) Examples of stable arches observed at a constriction of width $W/d=1.7$ with $N_{\rm p}=2$, $N_{\rm p}=3$, $N_{\rm p}=4$, and $N_{\rm p}=5$ particles respectively. Scale bars are $1\,{\rm mm}$. (b) Probability $p(N_{\rm p})$ to obtain an arch of $N_{\rm p}$ particles for a constriction $W/d=1.7$. For dilute and moderate solid fraction $\phi < 0.4$ (yellow histogram) only arches made of 2 particles were observed, whereas for $\phi \simeq \phi_{\rm max} =0.95$ (red histogram) arches made of more than 2 particles can be observed, although with a smaller probability.} 
  \label{fgr:figure6}
\end{figure}

As expected from the assumption of the stochastic model, large discrepancies between the experiments and the model are observed for dense suspensions, with a surface fraction close to the maximum packing fraction $\phi \simeq 0.95$. Such a situation can be encountered, for instance, when a fluid device has initially clogged so that particles accumulate at the constriction. When the clog is suddenly removed, this leads to a transient dense flow of particles. Figure \ref{fgr:figure5}(a) shows that the model fails to capture the dense case at $\phi \simeq 0.95$ (symbols at the right side of the figure). The estimated number of particles flowing through the constriction before clogging is much lower than what is predicted by the stochastic model. In this dense case, the system is already close to the maximum local packing fraction at the constriction. The quasi-bidimensionality of the channels allows the particles to move slightly out of plane to accommodate some additional densification without jamming and thus generate a distribution of avalanche sizes, as observed in quasi-bidimensional dry granular material.\cite{janda2008jamming}

A first issue with the dense suspensions is that the stochastic model does not account for possible overlapping of particles and hydrodynamic interactions between particles. This model is therefore not expected to be valid when the surface fraction of particles becomes too large. In addition, a significant difference between sufficiently dilute and dense suspensions of particles lies in the formation of arches. Indeed, we assumed previously that the number of particles required to clog the constriction corresponds to $n = \lfloor {W}/{D} \rfloor + 1$, for instance $n=2$ for $W/d=1.7$, and that the arch needs to be stable. However, arches of more than $n=2$ particles can be formed in this case for large particle surface fractions as illustrated in figure \ref{fgr:figure6}(a). Similar observations with dry grains in silos have recently reported the distribution of the number of particles in the arches and showed that, among other parameters, the size distribution of the arches depends on the angle of the constriction.\cite{lopez2019effect} For dilute suspensions, an analysis of all arches formed in our experiments revealed that only arches of $n$ particles are observed, whereas for dense suspensions various arch sizes $N_{\rm p} \geq n$ can be observed, as reported in figure \ref{fgr:figure6}(b). 

\begin{figure}[ht]
  \begin{center}
    \includegraphics[width=\linewidth]{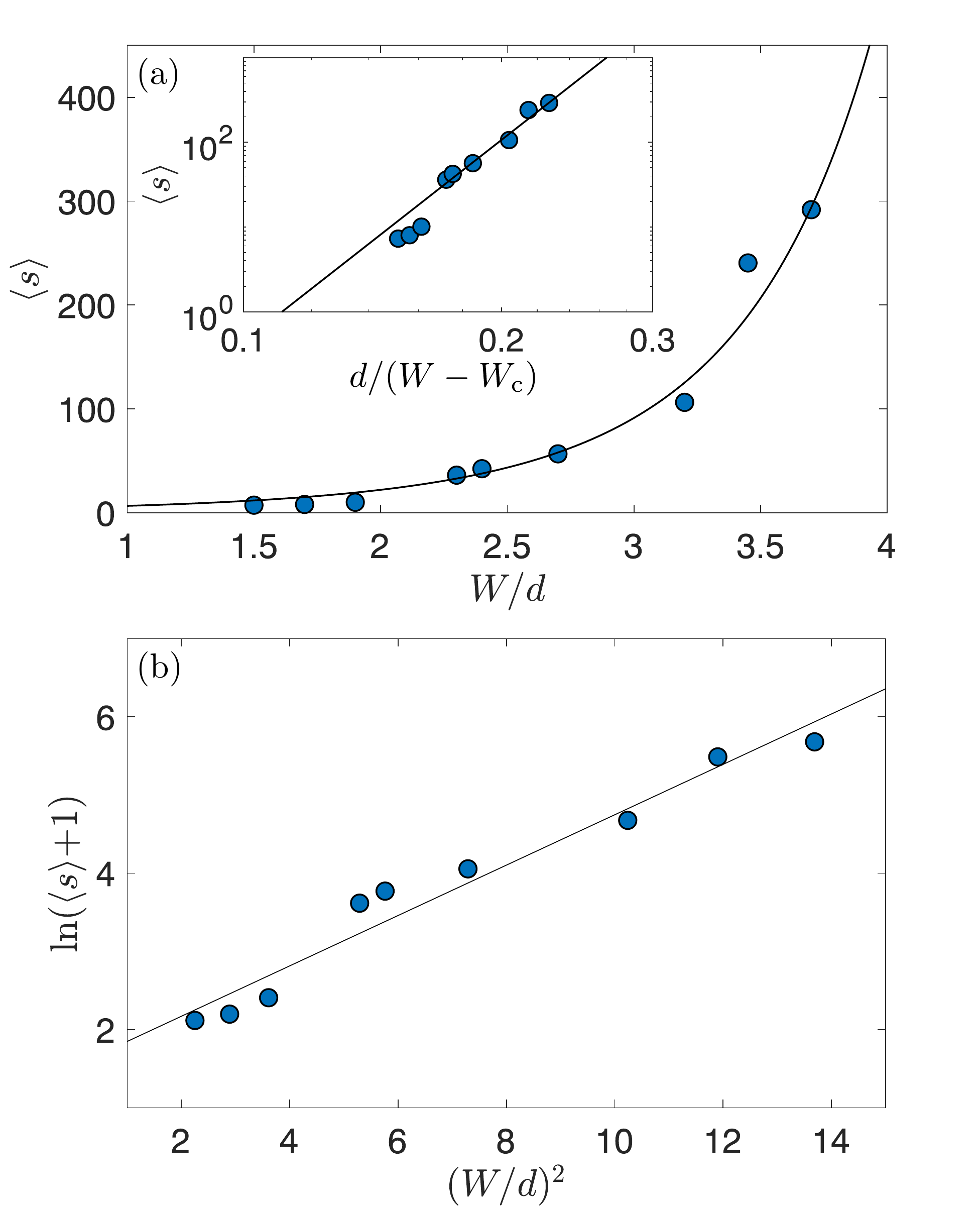}
  \end{center}
  \caption{(a) Evolution of the average number of particles flowing through the constriction before clogging $\langle s \rangle$ as a function of the rescaled width of the constriction $W/d$ for a dense suspension. The solid line shows a power-law divergence [Eq. (\ref{eq:powerlaw})]. Inset: Evolution of $\langle s \rangle$ as a function of the inverse of the distance to the fitted critical width $W_{\rm c}$. (b) Plot of $\ln(\langle s \rangle  + 1)$ as function of $(W/d)^2$, which is the signature of an exponential distribution. The solid line shows the best linear fit.} 
  \label{fgr:figure7}
\end{figure}

The situation with dense suspension, close to the maximum packing fraction, is actually much closer to the clogging of dry grains or immersed grains in silos than the stochastic model developed previously.\cite{zuriguel2005jamming,janda2008jamming,koivisto2017effect} In 2D systems with dry grains, the possibility of fitting the experimental data by either an exponential or a diverging power-law is still debated.\cite{janda2008jamming,zuriguel2014clogging} For our system in the dense regime, we report in Fig. \ref{fgr:figure7}(a) a power-law divergence of the average number of particles flowing through the constriction before clogging:\cite{janda2008jamming}
\begin{equation} \label{eq:powerlaw}
    \langle s \rangle= \frac{C}{(W_{\rm c}/d-W/d)^{\gamma}}.
\end{equation}
Our experimental results can be reasonably well captured by such an evolution with the fitting parameters $W_{\rm c}/d=8.1$, $\gamma=7.9$, and $C=3.7 \times 10^7$ (the goodness of the fit is illustrated in the inset of Fig. \ref{fgr:figure7}(a)). Those results are comparable to the ones found in the situation of 2D dry silos.\cite{janda2008jamming} However, in their system, Janda \textit{et al.} have reported that the evolution of $\langle s \rangle$ can be fitted by an exponential evolution:
\begin{equation} \label{eq:exponential}
    \langle s \rangle= A^{-1}\,{\rm exp}\left[B\,(\eta_0\,W/d)^2 \right]-1,
\end{equation}
where $A$, $B$, and $\eta_0$ are fitting parameters. We report in Fig. \ref{fgr:figure7}(b) the evolution of $\ln(\langle s \rangle  + 1)$ as function of $(W/d)^2$. In the range of parameters allowed by our experiments, the fit is as good as the one for a power-law so that we cannot establish the existence or not of the divergence. Indeed, the deviation from a power-law behavior to an exponential behavior for dry grains in 2D silos has been reported for $(W/d)^2>15$,\cite{janda2008jamming} which is beyond the range of our experimental data. 

Nevertheless, the main message is that for very dense suspensions with particles mostly in contact, the clogging of the constriction seems to follow the behavior observed in silos for dry and immersed granular materials. Therefore, the nature of the clogging events is very different to dilute suspensions, and to study the clogging behavior of suspensions, one needs to work below the maximum packing fraction.

\section{Conclusions}

Clogging of fluidic systems by suspensions of particles is a challenging process to consider and model, although important in many practical situations. Among the different clogging mechanisms, we focused in this study on the bridging of suspensions.  We experimentally investigated the influence of the solid fraction of non-Brownian particles on the clogging probability in a constricted millifluidic channel. Using a quasi-bidimensional experimental system, we were able to directly measure the local surface fraction of particles $\phi$ and the average number of particles $\langle s \rangle$ flowing through the constriction before a clogging event occurs. We showed that the solid fraction has a crucial role on $\langle s \rangle$, and that a constricted system will eventually clog for small enough width to particle diameter ratio $W/d<3$, even for small $\phi$. The number of particles $\langle s \rangle$, or alternatively the volume of suspension, flowing through the constriction before clogging becomes larger for smaller surface fraction. We also reported that larger values of $W/d$ increases drastically the duration of an experiments before a clogging event occurs, in agreement with a previous study performed in a 3D microfluidic system.\cite{marin2018clogging} 

The evolution of $\langle s \rangle$ when varying $\phi$ was rationalized assuming an initial random distribution of the particles in the channel upstream of the constriction and that particles follow passively the streamlines. We then showed that, to correctly predict $\langle s \rangle$, one has to consider that each particle arriving at the constriction can lead to the formation of an arch if $n-1$ particles, where $n=\lfloor W/d \rfloor +1$, are located in a small volume $\Sigma=(n-1)\,\pi\,d^2/(4\,\phi_{max})$ around this particles. The value of $n$ corresponds to the minimum number of particles that needs to span the constriction to form an arch. In addition, after the condition of forming an arch, one has to consider the probability that the arch is stable. This probability  decreases with the number of particles $n$ forming it.\cite{to2001jamming} Such an approach allowed us to capture the experimental evolution of the clogging probability when varying the solid fraction.

We should emphasized that although a stochastic model predicts correctly the evolution of $\langle s \rangle$ with $\phi$ and $W/d$, this model has several limitations. First, we assumed a random distribution of particles in the channel but hydrodynamics particle-particle interactions could result in a non-random distribution of particles. Furthermore, we used an empirical expression for the stability of the arch formed, and this expression is likely going to depend on the roughness of the particles,\cite{hsu2021roughness} the angle of the constriction, and other geometrical and dynamical parameters. Another difference is that although our experiments show that for non-concentrated suspension, only arches with $n$ particles were observed in all of our experiments, for dense suspensions close to the maximum packing fraction arches with more particles can also be observed. The clogging dynamics in this dense regime was similar to the one observed in dense dry grains in silos.   
 
Our study demonstrated that the solid fraction of a suspension of particles has a crucial influence on clogging by bridging. Even if a system may eventually clog for small enough width to particle diameter ratio $W/d$, one can optimize the concentration of the suspension to delay clogging in practical applications, such as in additive manufacturing.\cite{croom2021mechanics} Lastly, one must be careful when characterizing the clogging tendency of a system, as arch formation becomes increasingly rare with diluted suspensions but remains possible.

\section*{Acknowledgements}
This work benefited from the UCSB Academic Senate Faculty Grant, and partial support from the U.S.-Israel Binational Agricultural Research and Development Fund (BARD) US-5336-21. The authors thank J. Moehlis, A. Marin and H. Lhuissier for helpful discussions.

\balance

\bibliography{Clogging_Bridging} 
\bibliographystyle{ieeetr} 

\end{document}